\documentstyle[preprint,aps]{revtex}
\tightenlines
\begin{document}
\draft
\title
{\bf The first excited states of $^9$Be and $^9$B }
\author{V.D. Efros$^{1)}$ and J.M. Bang$^{2)}$}
\address{
1) Russian Research Centre "Kurchatov Institute", Kurchatov Square 1,
123182 Moscow, Russia\thanks{e--mail efros@polyn.kiae.su}\\
2) The Niels Bohr Institute, Blegdamsvej 17, DK--2100 Copenhagen, Denmark
\thanks{e--mail bang@nbivms.nbi.dk}}
\date{\today}
\maketitle
\begin{abstract}

It is found here that the $1/2^+$ first excited state of $^9$Be is a 
virtual 
state with the energy of -23.5 KeV. 
The line shape for the excitation of the state is
approximated with a simple analytic form based on the effective
range expansion. The partner in $^9$B of this state is found to be
a resonance with a maximum in the peak at about 1.1 MeV, FWHM 
of 1.5 MeV, and complex energy  of $0.6-i0.75$ MeV. 
The line shape for its 
excitation is calculated in terms of the
$p-^8$Be phase shift.
The phase shifts are obtained from 
$N-^8$Be effective potentials deduced 
from the data
on the photodisintegration of $^9$Be. A possibility for 
direct extraction
of the energy of the resonant state from experimental data 
is also discussed, and an expression for a residue at a 
virtual state pole in terms
of a quadrature taken over the 
virtual state eigenfunction is given. 

\end{abstract}
\bigskip
\pacs{PACS numbers: 
21.10.Pc, 23.50.+z, 24.30.Gd, 27.20.+n, 03.65.Nk\\}

\vfill\eject

\section{INTRODUCTION}
In the present work,  we obtain 
the properties of the first excited 
states of $A=9$ nuclei.
At low energy these nuclei provide a clean example of 
three--cluster systems with
not easily distortable constituents, thus serving as a test ground 
for theoretical multicluster
approaches, see e.g. \cite{varga,vor}. 
Reliable information on the properties of these nuclei would 
therefore be very valuable. However, extraction of this information
directly from the available experimental data is 
hampered by their ambiguities.
The corresponding data on the low--energy photodisintegration of 
$^9$Be are not
in mutual agreement, see \cite{eopt98}. All states in $^9$B
are particle unbound which hinders the search for excited
states.  
There exists a long--standing
controversy concerning the properties of the $^9$B 
first excited state, as obtained both 
experimentally
and theoretically, see \cite{bertsch85,exp}. Theoretical models
can help to analyze the data and also guide future experiments.
Along these lines, in Ref. \cite{eopt98} a semi--microscopic
model to describe the low--energy photodisintegration of $^9$Be
has been developed. An estimation of the reliability of
various data sets has been obtained with the help of 
this model, and a theoretical photodisintegration
cross section has been derived for astrophysical applications.
The model provides $N-^8$Be effective potentials that
reproduce the energy dependence of the cross section. Using them 
the properties 
of the first excited state of $^9$Be are extracted below 
and estimates for its
$^9$B partner are obtained.
  
In the next section we elucidate the nature of the 
first excited state of $^9$Be and
obtain its position. We obtain an analytic form for the
line shape of the state and investigate to what 
degree the line shape is independent
of the specific excitation process. In Sec. 3 
the first excited
state of the $^9$B nucleus is studied. 
The expression for the line shape is given, and
the position, width and the
complex energy of the state are calculated. A possibility to
obtain the latter quantity from a direct fit to the line shape is also
discussed.
In Sec. 4 our results are discussed along with those in the
literature.

Our considerations on the shape of the line in Sec. 2 and 3
are of a rather general applicability. 
In the Appendix a formula expressing a residue at a 
virtual state pole in terms
of a quadrature taken over the 
virtual state eigenfunction is given. 

\section{THE FIRST EXCITED LEVEL OF $^9$BE}
We proceed from the dynamic input for the description of the
system obtained in Ref. \cite{eopt98}.
In that work, the $^9$Be photodisintegration 
cross section has been 
calculated
in the framework of the following model. 
The three--body $\alpha+\alpha+n$
representation of
the system has been adopted. The $^9$Be ground state  
wave function has been 
calculated from the three--body dynamic equation with 
$\alpha\alpha$  
and $\alpha n$ potentials. The final continuum state 
has been chosen as a
 product of 
the intrinsic wave function of the $^8$Be resonance 
and the $n-^8$Be 
relative motion function.
The latter wave function has been calculated from the 
Woods--Saxon potential
whose parameters were determined by fitting to several 
radioactive isotope
data.

The model thus gives $n-^8$Be effective potentials that
allow extracting the properties of the first excited
state of $^9$Be. First, let us comment on the status of the
results obtained in this way. 
The model accounts for only the two--body
$n-^8$Be photodisintegration and disregards the 
direct
three--body $\alpha+\alpha+ n$ disintegration. 
The cross section for the direct
$\alpha+\alpha+ n$ disintegration  
is presumably very small due to the 
threshold regime which is confirmed in experiment, see \cite{eopt98}.
Further, the properties of the state considered are determined by
the phase shift $\delta$ of $n-^8$Be scattering. So, the question arises
whether this phase shift can be correctly obtained with the help 
of the model. In this connection one can admit that
in the vicinity of the first 
excited  state of $^9$Be
the energy dependence
of the photodisintegration cross section is 
also determined by this phase shift. To a degree this holds true, 
use of the two--body $^8$Be$+n$ 
dynamics to extract
the properties of the state does not contain any restrictions.
In Ref. \cite{eopt98} the cross section has been fitted for 
excitation energies up to 0.5 MeV, and in this region 
the corresponding energy
dependence $\sin^2\delta/k$ dominates the cross section, see Fig. 1 below.
Moreover, the initial $^9$Be ground state is described realistically
in this model.
Therefore, use of the two--body dynamics seems to be sufficient for 
extracting
the properties of the state considered. In general, these properties
can also be extracted via the fit of an assumed line shape
directly to the data. However, use of the two--body dynamics in conjunction
with shell--model considerations allowed selection between alternative 
data sets in Ref. \cite{eopt98}. Below it will also help us to
obtain the properties of the analog $^9$B state relying on
the $^9$Be photodisintegration data.

The fit to the photodisintegration cross section \cite{eopt98} 
provides the potentials
\begin{equation}
V(r)=V_0 /[1+e^{(r-R)/b}] \label{ws}
\end{equation}                                             
describing $n-^8$Be scattering. The best version
\begin{equation}                                                               
V_0=35.99\,\,{\rm MeV},\,\,\,\,\,\,  R=3.126\,\,{\rm fm}, 
\,\,\,\,\,\,
b=0.8108\,\,{\rm fm} \label{par}           
\end{equation}
will be used below. Other potential versions lead to similar 
results as 
commented below.

Let us denote $E=(\hbar k)^2/(2\mu)$ the energy of the relative 
$n-^8$Be motion.
The potential (\ref{ws}), (\ref{par}) leads
to a  $n-^8$Be scattering amplitude  $f(k)$ having a pole  
at $k=-i\kappa$, $\kappa>0$:
\begin{equation}
f(k)\rightarrow \frac{ic_0}{k+i\kappa} \label{f}
\end{equation}
as $k\rightarrow -i\kappa$. This means 
that the first excited state of $^9$Be is
a virtual state strongly
 coupled to the
$n-^8$Be channel. The energy of the state is
\begin{equation}
-(\hbar \kappa)^2/(2\mu)=-\bar{E}=-23.53\,\,{\rm KeV}. \label{e}
\end{equation}
This value was obtained via solving the 
eigenvalue problem with the virtual state boundary condition, 
cf. the Appendix.

Let us consider the shape of the line for a transition proceeding via 
the virtual state. The shape depends partly on the 
specific way 
of excitation of the state. The main energy dependence
is universal, however,  and it is determined by the 
intrinsic properties
of the state. We shall compare the universal energy dependence to the
energy dependence for the process of $^9$Be photodisintegration and thus
estimate the dependence of the line shape on the specific process.

We assume that the cross section for formation of 
$^9$Be in the continuum 
with the energy $E$ in the
vicinity of the excited level
can approximately be presented in the form
\begin{equation}
\sigma\sim\int k^2dk|\langle\Phi|\Psi_k^-(J=1/2^+)\rangle|^2\delta
 (k^2/(2\mu )-E). \label{crs}
\end{equation}
Here $\Phi$ is some localized state in the subspace of the $^9$Be 
degrees of freedom. It includes  the g.s. of $^9$Be and the 
(effective) transition operator accounting for the excitation 
mechanism (photodisintegration, e.g.). The transition operator may 
have an energy dependence that is smooth at 
$E\rightarrow 0$.
The function $\Psi_k^-$ is the continuum spectrum function that
corresponds to 
outgoing  $n$ and $^8$Be  fragments in the relative $s$--state 
thus describing 
the two--body 
$^8$Be$+n$ photodisintegration.
The representation 
(\ref{crs}) implies
that the formation process can be described in the subspace of the $^9$Be
degrees of freedom. Under this condition properties of a state 
reveal themselves independently of the interactions with other particles
participating in its formation.\footnote{In the framework
of the model \cite{eopt98} 
only the two--body $^8$Be$+n$ channel is retained
in the function $\Psi_k^-$ while 
the incoming three--body $\alpha +\alpha +n$ channels
are disregarded.   
This model assumption  is also inherent to
all the previous work. One can see, however, that 
to a first approximation this
does not change the energy dependence of the cross section and 
thus does not influence the results.}
In Eq. (\ref{crs}) the normalization condition 
$\langle\Psi_k^-|\Psi_{k'}^-\rangle=\delta(k-k')/k^2$
should be fulfilled, so that the $n-^8{\rm Be}$ relative 
motion function in $\Psi_k^-$ at large 
distances is normalized to $e^{-i\delta}\sin(kr+\delta)/(kr)$. 

We may write the $k$--integral of (5) as a contour integral where the pole 
contribution is given by (3), neglecting influence of other degrees of 
freedom in $\Phi, \Psi^-_k$.

The energy dependence of the cross section (\ref{crs}) in 
the vicinity of the virtual
level is $\sin^2\delta/k$. This energy dependence arises
if one replaces the outer part
$\sin(kr+\delta)/(kr)$ of the $n-^8$Be relative motion function 
in the matrix
element from 
Eq. (\ref{crs}) by $\sin\delta/(kr)$. This can be done under 
the conditions $(kR)^2
\ll 1$, and $R\ll |a|$. Here $a$ is the scattering length, and $R$ 
is chosen such that the relative contribution 
of $r>R$ values 
to Eq. (\ref{crs}) is small.  $R$ exceeds the range of the $n-^8$Be 
interaction, and
the inner part
of the final state wave function $\Psi_k^-$ acquires the same energy 
dependence as the 
outer one. 
The $\sin^2\delta/k$ energy dependence 
is nothing else as the so called Migdal--Watson factor \cite{lan77}.

We can rewrite the corresponding contribution to the 
cross section as const$\cdot k|f(k)|^2$ where
$f(k)$  is the $s$-wave
$n-^8$Be scattering amplitude. Almost the same accuracy is 
kept if one takes 
$f(k)$ within the effective range approximation
\begin{equation}
f(k)=[-1/a+(1/2)r_0k^2-ik]^{-1}. \label{f1}
\end{equation}
Then one obtains
$|f(k)|^2\sim [(E+\bar{E})(E+E_1)]^{-1}$. In the 
model (\ref{ws}), (\ref{par}) $E_1=1.569$  MeV.
Using this expression, 
it is convenient to present the cross section in the following form
\begin{equation}
 \sigma (E)= \sigma _m\frac{2\sqrt{E\bar{E}}}{E+\bar{E}}
\frac{\bar{E}+E_1}{E+E_1}. \label{sig}
\end{equation}
Since  $\bar{E} \ll E_1$ the energy dependence of the cross section 
from the threshold 
up to its maximum is entirely determined by the second factor 
in Eq. (\ref{sig}).
It is then seen that the maximum of the cross section occurs 
practically at 
the $\bar{E}$ value. With a high precision $\sigma_m$ is the value 
of the cross section 
at the maximum. As one can see from Eq. (\ref{crs})
the lowest order correction to   Eq. (\ref{sig})
is of the form $1-(E/E_2)$ where $E_2$  
is not small and depends on a specific excitation process. To a certain 
degree it can be accounted for via a renormalization of 
$E_1$. Eq. (\ref{crs})
is the Breit-Wigner formula with $\Gamma=\Gamma(E)$,
(or, the one-level $R$ matrix expression) rewritten in a different form.

In Fig. 1 various universal expressions for the shape of the 
line are compared with the exact line shape for the photodisintegration
process calculated in the model of Ref. \cite{eopt98}. 
($E=E_\gamma-E_{th}$.) 
The full curve represents the
latter line shape, the long--dashed curve is the energy dependence 
$\sim k|f(k)|^2$,
the dash-dotted curve
represents the expression (\ref{sig}), and the dotted curve is the energy 
dependence given
by the second factor from Eq. (\ref{sig}).
The FWHM value for 
the photodisintegration process provided by the first mentioned 
curve is 196 KeV. 
The FWHM
values provided by the next two curves are 230 and 240 KeV. 
The latter two values are process--independent.
Thus FWHM may depend rather sizably on the excitation process. On the 
contrary, the  position of the maximum in the peak is practically 
process--independent 
coinciding with the absolute value of the energy of the virtual 
state. 

Experimentally, the energy dependence $E^{1/2}(E+\bar{E})^{-1}$, 
specific
to a virtual state could be confirmed
by measuring the shape of the line 
from the threshold up to the region of the maximum with the 
tagged photon techniques. A
simpler, while indirect, way is to study the whole peak in the $(e,e')$
reaction and extract the properties of the state in the way similar to that
used above for photodisintegration. In case of
accurate and detailed $(e,e')$ or $(p,p')$ data the energy $\bar{E}$ 
of the state could
also be extracted from the fit of the Eq. (\ref{sig}) form or its 
above--mentioned extension.

Besides the energy $\bar{E}$, or the pole position of the 
$n-^8$Be scattering
amplitude, another quantity to be reproduced in a microscopic 
calculation is
the residue $c_0$ in the pole, Eq. (\ref{f}). The exact $c_0$ value 
was calculated with
the formula derived in  the Appendix that 
represents  $c_0$ as an integral taken over the virtual state eigenfunction.
The result is $c_0=0.7837$. 
We note that with high accuracy both $\bar{E}$ (or $\kappa$, see (\ref{e}),)
and $c_0$ 
can   be expressed in terms of the  $n-^8$Be scattering 
length 
$a$ and the 
effective range $r_0$ . From 
Eq. (\ref{f1}) one obtains
\begin{eqnarray}
\kappa=r_0^{-1}[(1-2r_0/a)^{1/2}-1], \nonumber\\
c_0=(1-2r_0/a)^{-1/2}=(1+\kappa r_0)^{-1}. 
\end{eqnarray}
The effective range parameters for the potential (\ref{ws}), 
(\ref{par}) are 
\begin{equation}
a=-27.65\,\, {\rm fm} ,\,\,\,\,\,\,\,\,\,\,\,\, r_0=8.788\,\, {\rm fm}, 
\label{les}
\end{equation}
and the expected accuracy of the above expressions is  about 
$|r_0/a|^3\simeq 1\%$. Their numerical values proved to be
even more accurate: $\bar{E}=23.51$ KeV, $c_0=0.7819$.
As it is explained in \cite{eopt98} all other
acceptable potentials found there lead to the $a$ and $r_0$ values 
very close to those 
in Eq. (\ref{les}). Hence the properties of the virtual 
state given by these potentials are quite similar.

\section{THE FIRST EXCITED LEVEL OF $^9$B}
To calculate the resonant peak for an excitation of the $^9$B $1/2^+$ 
level, we proceed 
again from Eq. (\ref{crs}), only with a different expression 
for the pole term but still
assuming that there is only one decay channel, $p-^8$Be, for the state 
considered.  
In the $^9$B case, 
disregarding the possibility for the decay into 
the $N+\alpha+\alpha$ channel, being less substantiated 
than in the $^9$Be case
because 
of a higher energy with respect to the three--body threshold,
seems still to be reasonable. The $n-^8$Be potentials obtained 
in Ref. \cite{eopt98} lead
presumably to a good description of the $n-^8$Be $1/2^+$ phase 
shifts. Then one may
suggest that the $p-^8$Be $1/2^+$ phase shifts will also be 
properly reproduced 
with these
potentials with an addition of the Coulomb interaction. This will 
suffice to obtain approximately the main
properties of the $^9$B $1/2^+$ level. 

Beyond the range of the nuclear potential
the $p-^8$Be relative motion function $\chi(r)/kr$ 
entering $\Psi_k$ in Eq. (\ref{crs})
may be represented
in the form
\begin{equation}
\chi=G\sin\delta +F\cos\delta =(\sin\delta/C)[C(G+F\cot\delta)]
\equiv(\sin\delta/C)\phi(k,r). \label{ph}
\end{equation}
Here $\delta$ is the nuclear phase shift, 
$C^2=2\pi\eta[\exp(2\pi\eta)-1]^{-1}$ is the Coulomb 
penetrability factor, and $F$ and $G$ are the Coulomb functions. The possible resonant 
pole is contained in
the factor $\sin\delta$ while the function $\phi$ 
defined in Eq. (\ref{ph}) is smooth at
low energies. (At $k\rightarrow 0$ 
$\phi(k,r)\rightarrow zK_1(z)-(a_c/4a)zI_1(z)$,
$z=2(2r/a_c)^{1/2}$, where $a$ is the scattering length, and $a_c$ 
is the Bohr radius.)
To a first approximation one may use $\phi(k_0,r)$
 in the calculation 
of the cross section (\ref{crs}),
$k_0$ being chosen in the vicinity of the resonance. Hence 
the energy dependence of the 
excitation cross section in the resonant peak region 
is given by the expression 
\begin{equation}
\frac{\sin^2\delta}{kC^2}=k\frac{|f(k)|^2}{C^2}=
\frac{kC^2}{(kC^2\cot\delta)^2+(kC^2)^2}.
\label{res}
\end{equation}
Here $f$ is the nuclear
amplitude of $p-^8$Be scattering.  Thus, quite naturally, 
up to process--dependent corrections the resonant
cross section for the excitation of the state is proportional to 
the scattering cross 
section times a universal factor  
proportional to a width. 
The quantity $kC^2\cot\delta$ in (\ref{res}) allows the 
well--known representation
$[-1/a+(1/2)r_0k^2+\ldots]-(2/a_c)h(1/a_ck)$, showing that it 
is smooth and finite at $E=0$. The $kC^2$ behavior of the cross section at 
$E\rightarrow 0$ is seen directly from Eq. (\ref{crs}) and the properties 
of the
Coulomb functions.

The $p-^8$Be phase shifts entering Eq. (\ref{res}) were obtained from the Schr\"odinger equation 
with the nuclear
potential (\ref{ws}), (\ref{par}) plus the Coulomb interaction. The 
latter took into account
the density distribution in $^8$Be and the charge distribution in 
the $\alpha$ particles:
\begin{equation}
V_{Coul}(r)=(4e^2/r)(2/\pi)\int_0^\infty(\sin qr/q)F_\alpha(q)I(q)dq,
\end{equation}
where $F_\alpha(q)$ is the charge form factor of the $\alpha$ 
particle \cite{ott}, and 
\[ I(q)=\int_0^\infty (2/q\rho)\sin(q\rho/2)\psi^2(\rho)d\rho, \]
$\psi$ being the $^8$Be wave function.

The energy dependence (\ref{res}) obtained indeed proved to exhibit a 
pronounced peak, 
and it is shown in Fig. 2 with 
a solid curve. The position $E_{max}$ of the maximum in the peak with 
respect to 
the $p-^8$Be 
threshold and the FWHM value are the following: $E_{max}=$1.13 MeV, FWHM=1.64 MeV. 
The long--dashed curve in Fig. 2
represents the peak for the other acceptable nuclear potential of the 
Woods--Saxon form  found in Ref. \cite{eopt98} whose parameters are
\begin{equation}                                                               
V_0=52.86\,\,{\rm MeV},\,\,\,\,\,\,  R=2.006\,\,{\rm fm}, \,\,\,\,\,\,
b=1.051\,\,{\rm fm}. \label{p2}           
\end{equation}
For this case $E_{max}=1.02$ MeV, and FWHM=1.43 MeV. 

If the peak obtained corresponds to a state then the scattering
amplitude should have a pole in the vicinity of $E_{max}$. Thus
we shall search for such a pole.
One can show that when the long--range
Coulomb interaction is switched on, virtual states cease to exist, 
turning normally 
to complex--energy resonances, and this applies to our case.
(It applies also to, e.g., $p--p$ scattering where, in contrast to
what is often said, there are no virtual states.) 
The energy $E=E_0-i\Gamma/2$ of a resonance  being connected to the pole of 
a scattering amplitude, 
is process--independent and thus characterizes a resonance quite precisely 
even in case of a broad width. 

In our case this quantity is computed as a complex eigenvalue 
in the $p-^8$Be Schr\"odinger equation. Here we 
used the codes of Ref. \cite{thompson} or, when 
this is inoperative, the $\rho$--series (14.1.4) for $F$ and
the corresponding expansion (14.1.14) -- (14.1.19) for $G$ from
Ref. \cite{abram}.\footnote{We rewrite the latter expansion for
$l=0$ in the form valid for complex $\eta$ values:
\[ G(\eta,\rho)=C^{-1}\{2\eta\rho\Phi(\eta,\rho)
[\ln(2\rho)+(1/2)[\psi(i\eta)+\psi(-i\eta)]+2\gamma-1]+
\sum_{k=0}^\infty a_k(\eta)\rho^k\}. \]
Here $\psi(z)=\Gamma'(z)/\Gamma(z)$, $\gamma$ is the 
Euler constant, and the quantities $\Phi(\eta,\rho)$ and $a_k(\eta)$
are defined in \cite{abram}.} These expansions are fast convergent for not
extremely high values of $|\rho|$, $|\eta|$.

The resonant pole was found with the following parameters: 
$E_0=0.60$ MeV, 
$\Gamma/2=0.77$ MeV, and $E_0=0.56$ MeV, $\Gamma/2=0.70$ MeV for 
the potentials 
(\ref{par}) and (\ref{p2}), respectively. The positions $E_0$ of the 
resonance are
shifted downwards with respect to the maxima in the peak found above, 
while the widths $\Gamma$
are close to the FWHM values. Similar trends were observed for 
$^5$He, $^5$Li 
resonances \cite{eo96}. It is interesting to note that the height 
of the Coulomb
barrier in our case proved to be 0.63 MeV only, i.e. the resonance
with a finite width
sits at the top of the barrier. This is possible only when the
width of the resonance becomes so broad that it is comparable to $E_0$.

In conclusion, let us add the following two comments. First,
since the energy $E=E_0-i\Gamma/2$ of the resonance is a
convenient process--independent quantity to compare with theory, 
it would be very useful to extract it directly 
from future experimental data without constructing a model 
for the excitation
process. Such a task has been accomplished successfully in some cases,
see e.g. \cite{eo96,hale87}. However it is not clear whether this is possible
in practice for the broad resonance we consider. We perform a numerical
experiment to clarify this issue. We  explore the 
possibility to reconstruct the $E_0$ and $\Gamma$ values obtained 
proceeding from a reasonable fit to the resonant peak of the 
form (\ref{res}), see Fig. 2, calculated in the same
model, Eq. (\ref{par}). We fit the peak with the expression
\begin{equation}
a_1\frac{kC^2}{[a_2+a_3(E-E')+a_4(E-E')^2]^2 +(kC^2)^2} \label{fit}
\end{equation} 
related to Eq. (\ref{res}). Here $E'=1.5$ MeV, the fit is extended
over the values $(E'-\Gamma/2)\le E\le (E'+\Gamma/2)$ at which the
cross section exceeds a half of its value at the maximum, and $a_i$
are fitting parameters. At choosing the form of the expression 
(\ref{fit}) it was taken into account that the effective range approximation
is not accurate enough in the peak region and that $kC^2\cot\delta$ 
has no minimum in that region in our case. The resonant $E$ value, 
$E=E_0-i\Gamma/2$, 
is obtained as a zero of the expression 
$a_2+a_3(E-E')+a_4(E-E')^2-ikC^2$ that corresponds to the pole of the 
scattering amplitude.

The search of the least square
minimum was performed with respect to $E_0$, $\Gamma$, and $a_3$ 
with values taken on some grids. At given values of these parameters
$a_2$ and $a_4$ were fixed via equating to zero real
and imaginary parts of the above expression. The $a_1$ value was found
analytically from the least square minimum requirement. (If 
$kC^2\cot\delta$ is completely reproduced by the fit, we should
have $a_1=1$.)

It occurred that the fit is unstable in this form, and $a_1$ takes
unrealistic values. Let us then remove the search of the $\Gamma$
value admitting instead the hypothesis $\Gamma=$FWHM. This 
condition is fulfilled with  sufficient accuracy in our case, and also
in the $^5$He, $^5$Li cases  \cite{eo96}. This modified procedure
leads to $E_0$ values that are quite stable and close to the true one.
The stability was checked by changing the stepsize in the $a_3$
parameter. 

The second comment concerns the value of the effective range 
$r_0$ for the $p-^8$Be 
scattering. For the potential (\ref{par}), for example, it 
proved to be 3.01 fm i.e.
much smaller then that from Eq. (\ref{les}) for the $n-^8$Be case. 
This, however,
does not mean that the range of the $p-^8$Be nuclear interaction 
is sizably
different from that for the $n-^8$Be interaction. The effective 
range can serve as a
measure of the interaction range when $a\gg r_0$ and, in addition, 
the zero energy 
scattering wave function $u(r)$ entering the effective range 
definition is close to unity at
the edge of the well $R$. This holds true for the neutral case 
but not when the
rather strong Coulomb interaction is present. The latter 
interaction suppresses the wave
function at the $R$ value, so that the range of nuclear 
interaction may be 
estimated as $r_0/u^2(R)$.     
  
\section{DISCUSSION}
Basing on the model of Ref. \cite{eopt98} we obtained that the first 
excited level
of $^9$Be is a virtual state with the energy $-\bar{E}=$-23.5 KeV with respect 
to the $n-^9$Be threshold. It is shown that the peak position for an excitation
of this state  practically coincides with the $\bar{E}$ value while the FWHM 
value of
the peak may sizably depend on a specific excitation process. Within the 
effective range 
approximation the properties of the state are reproduced with a high accuracy.
The line shape is
aproximated with a simple analytic form based on the effective
range expansion. 

In 
Ref. \cite{barker83} a single--level $R$-matrix fit to the 
photodisintegration data of Ref. \cite{fush82} led to the description
of the state considered as a
complex--energy resonant state. However the data \cite{fush82} are at variance 
with the
earlier data, and it was concluded in Ref. \cite{eopt98} that the latter ones
are preferable. The positions of maxima in the available experimental 
$^9$Be$(p,p')$ 
spectra \cite{tucker70,kuechler87}
do not contradict the above listed value of 23.5 KeV. 

In Ref. \cite{kuechler87}
the quantities pertaining to the $1/2^+$ state of $^9$Be were presented in the form 
of parameters
entering the Breit--Wigner formula
\begin{equation}
\sigma(E)=\frac{8\pi^2}{9}\frac{e^2}{\hbar c}E_\gamma B(E1,E_\gamma)
\frac{\Gamma(E)/2}
{(E-E_R)^2+[\Gamma(E)/2]^2}, \label{bw}
\end{equation}
that was fitted to the $(e,e')$ data of that paper. Here $\Gamma(E)=G\sqrt{E}$, 
$G$ being the reduced width of the level. 
The values $E_R=19\pm 7$ KeV, $\Gamma=217\pm 10$ KeV were reported, the latter value
refers presumably to $\Gamma(E_R)$. These values are quoted in the review article \cite{ajz88}.
They cannot be correct since they lead to a quite
unrealistic $\bar{E}$ value, i.e. that of the maximum of an excitation cross 
section, of 0.6 KeV. 
Based on Fig. 6 
from Ref. \cite{kuechler87} one may suggest, however,  that these values 
do not 
refer to $E_R$ and $\Gamma$
from Eq. (\ref{bw}) but to the position 
of the maximum of the cross section and the FWHM, respectively. If it is 
the case, 
these values are in agreement with ours.
(Absence of the information on energy dependence of $B(E1,E_\gamma)$ 
inhibits a precise 
conclusion concerning the FWHM.)

In Ref. \cite{varga} a microscopic study of the spectrum of $^9$Be 
at the 9--nucleon 
level was undertaken. The $1/2^+$ level was not detected at all.
The reason may lie in that the method of complex scaling used
in that work is suited for search of complex--energy resonant states but not 
virtual states.

Further, we studied the first excited state in
$^9$B in the present work using the nuclear potentials derived in 
\cite{eopt98} from the
analysis of the $^9$Be photodisintegration data and adding the Coulomb
interaction between $p$ and $^8$Be. An approximate 
process--independent expression for the line shape
of the state in terms of the
$p-^8$Be phase shift
is obtained. The position of the 
peak with respect to 
the $p-^8$Be threshold and its FWHM 
given by this expression are about 1.1 MeV and 1.5 MeV, 
respectively. The peak considered is
caused by a complex--energy pole in the $p-^8$Be scattering 
amplitude. The pole  
proved to be located
at energy $E=E_0-i\Gamma/2$ with $E_0\simeq 0.6$ MeV and  
$\Gamma\simeq 1.5$ MeV.

In Ref. \cite{bertsch85} a prediction for the peak of the $^9$B resonance 
was obtained: $E_{max}=1.13$ MeV, FWHM=1.40 MeV. It is close
to ours while the underlying assumptions of that work were rather different. 
Our common features with that work 
consist in use of $p-^8$Be dynamics to obtain the resonance and in 
obtaining the $p-^8$Be Woods--Saxon
potential from $^9$Be photodisintegration data. The big differences consist 
in the conditions from which 
parameters of the  potential are deduced, in the 
parameters themselves, and in a 
prescription to calculate
the resonance. In Ref. \cite{bertsch85} the two--body dynamics were 
used both for the inital ground
state and the final continuum state of $^9$Be. Depths of the potentials 
were varied whereas their range
and diffuseness were kept at their "classic" vaues. The data of 
Ref. \cite{fush82} were fitted (probably up to
a normalization). The fit was only moderately good. The line shape of 
the $^9$B resonance was computed as if
it were excited due to a fictitious dipole transition from the ground 
state of $^9$Be. Our potentials
were deduced \cite{eopt98} at the assumption of three-body 
$\alpha+\alpha+n$ dynamics for the ground state of
$^9$Be and two--body dynamics for its continuum. Range and 
diffuseness of the potential were varied
in addition to the depth. The data of  Ref. \cite{fush82} were 
concluded to be less preferable, and the alternative earlier 
data were fitted without freedom
in an absolute normalization. The fit is statistically quite 
good. The line shape of the $^9$B resonance was 
computed from Eq. (\ref{res}). The Coulomb interaction was treated 
more accurately, and also the pole position 
$E_0-i\Gamma/2$ was calculated.  
Keeping range and diffuseness of the potential
at their "classic" values was perhaps an important point
in the analysis of Ref. \cite{bertsch85} 
which allowed obtaining correct results 
even at assumptions that are not completely
valid. An interesting point is that the different 
prescriptions for calculating a resonance 
in our work and in Ref. \cite{bertsch85} lead to similar 
results. To check this we calculated the resonance
for the potential of Ref. \cite{bertsch85} with the prescription of
Eq. (\ref{res}). The results are $E_{max}=1.06$ MeV, FWHM=1.47 MeV, 
and they are close to those
reported in Ref. \cite{bertsch85}. (The difference in treatment 
of the Coulomb interaction should also be taken into account here.)

In Ref. \cite{barker87} the energy of the $1/2^+$ first 
excited state of $^9$B was derived from 
$R$--matrix parameters fitted  \cite{barker83} to the 
$^9$Be data of Ref. \cite{fush82} and from  
values of the Coulomb displacement energy calculated 
with the help of the shell model. The result $E_0\simeq 2$ MeV
differs considerably from that of our work and 
Ref. \cite{bertsch85} leading to the inverted value
of the so called Thomas--Ehrmann shift. As it is mentioned above 
in connection with our comparisons 
to Refs. \cite{barker83,bertsch85}, use of this 
particular set of data to derive
$R$--matrix parameters might be a disadvantage. 

We are indebted to M.V. Zhukov for useful discussion
in the course of this study.
A part of this work was done during the stay of V.D.E. 
at the Niels Bohr Institute, and he expresses his 
gratitude for the kind hospitality. The work was partially 
supported by Russian Foundation for Basic Research
(grants 96-15-96548 and 97-02-17003). 

\appendix
\section*{RESIDUE PROPERTIES OF VIRTUAL STATE POLES
\protect\\ OF A SCATTERING AMPLITUDE}
In the vicinity of the virtual state pole the two--body 
scattering amplitude behaves according to Eq. (\ref{f}).
We
obtain a formula that expresses a residue $c_0$ in terms of 
a quadrature taken over the virtual state eigenfunction. 
The result is 
$c_0=(-1)^l/I_v$, where $I_v$ is given by Eq. (\ref{int}) 
below with $\Delta(r)$ entering
the virtual state eigenfunction (\ref{del}). The latter function is
readily calculated from the Schr\"odinger equation.
(The consideration
is formally applicable to a motion in a central potential with any orbital 
momentum $l$ 
although the $l>0$ virtual state case is of 
little interest.)      

Let $\psi(r)$ be the solution  to the Schr\"odinger equation  
regular at the origin.  
Beyond the range of a potential 
the function $\chi(r)=r\psi(r)$ behaves as 
$\exp(ikr)+(-1)^{l+1}S^{-1}(k)\exp(-ikr)$. Complex $k$ values are allowed.
A virtual state corresponds to a pole 
of the $S$--matrix $S(k)$ at $k=-i\kappa$,
$\kappa$ being real and positive.\footnote{It is implied that at 
large $r$ the
potential decreases more rapidly than $\exp(-\kappa r)$, 
cf \cite{lan77}.} 
In terms of $\kappa$, 
$\chi(r)=\exp(\kappa r)+(-1)^{l+1}S^{-1}\exp(-\kappa r)$ at large $r$,
so the 
boundary condition for virtual states consists in absence of 
the decreasing
exponential. If the eigenvalue is sufficiently small 
then the eigensolution can be found
directly using a logarithmic derivative at sufficiently large $r$. This 
holds true in our
case. 
 
For bound states, the well--known formula corresponding to that we
want to derive is the following \cite{kra,hei}.
Let the bound state wave function be normalized to unity, 
and $A$ the coefficient in its asymptotics $A\exp(-\kappa r)$. 
Then
\begin{equation}
f(k)\rightarrow \frac {i(-1)^lA^2}{2\kappa}\frac{1}{k-i\kappa}  
\label{eq:poleb}
\end{equation}
as $k$ tends to $i\kappa$. 

In the virtual state case we shall
proceed from the relation
\begin{equation}
\frac{\partial \chi}{\partial r}\frac{\partial \chi}{\partial E}
-\chi\frac{\partial^2 \chi}
{\partial r\partial E}=\frac{2m}{\hbar^2}\int_0^r \chi^2(r')dr'. 
\label{eq:rel}
\end{equation}
It is similar to that used in Ref. \cite{lan77} for the derivation of Eq. 
(\ref{eq:poleb}). 
We take $r$ to be 
large enough so that in the vicinity of the virtual state pole 
$\chi(r,E)\simeq \exp(\kappa r)+\alpha(E+\bar{E})\exp(-\kappa r)$. 
Here $\bar{E}=(\hbar\kappa)^2/2m$. The pole
term in the scattering amplitude is then 
$(-1)^{l+1}(2\kappa\alpha)^{-1}(E+\bar{E})^{-1}$,
and the constant $\alpha$ is to be found. We denote $\chi_v(r)$ the
virtual state eigenfunction, so that $\chi_v=\chi(E=-\bar{E})$.
We set
\begin{equation}
\chi_v(r')=\exp(\kappa r')+\Delta(r'). \label{del}
\end{equation}
Here $\Delta(r)$ is a rapidly decreasing function.
We substitute Eq. (\ref{del})   
into the right--hand side of Eq. (\ref{eq:rel})
and equate the terms of the orders of $\exp(\kappa r)$ and unity  
in both sides of Eq. (\ref{eq:rel}) at $E$ tending to $-\bar{E}$. 
Then we obtain 
$2\kappa\alpha=-(2m/\hbar^2)(I_v/2\kappa)$, with
\begin{equation}
I_v=1-2\kappa\int_0^{\infty}[\Delta^2(r)+2\Delta(r)\exp(\kappa r)]dr. \label{int}
\end{equation}     
The product $\Delta(r)\exp(\kappa r)$ is a rapidly decreasing function, so
the integral converges. Finally,
\begin{equation}
f(k)\rightarrow \frac{i(-1)^{l}}{I_v}\frac{1}{k+i\kappa}  \label{eq:polev}
\end{equation}
as $k$ tends to $-i\kappa$, $\kappa>0$. 

To display an analogy between the bound and virtual state pole 
cases, we note
that the right--hand side of Eq. (\ref{eq:poleb}) can be rewritten as
\begin{equation}
\frac{i(-1)^l}{2\kappa N_b}\frac{1}{k-i\kappa}, 
\end{equation} 
where $N_b$ is the norm of the eigenfunction $\chi_b$ 
behaving as  
$\exp(-\kappa r)$ at $r$ tending to infinity. If one sets 
$\chi_b(r)=\exp(-\kappa r)+\Delta(r)$ then $2\kappa N_b$ takes 
just the form of
Eq. (\ref{int}) with the replacement $\kappa\rightarrow -\kappa$.

\begin{figure}
\caption{Line shapes for transitions to the first 
excited state of $^9$Be.
The solid curve represents the exact line shape for 
the photodisintegration process 
calculated with the model of Ref. [5]. The long--dashed 
curve is the energy dependence
$k|f(k)|^2$, where $f$ is the $n-^8$Be scattering amplitude. 
The dash--dotted curve 
stands for
Eq. (\protect\ref{sig}). The dotted curve represents the 
energy dependence given
by the second factor from Eq. (\protect\ref{sig}).}   
\end{figure}

\begin{figure}
\caption{Line shapes for transitions to the first 
excited state of $^9$B calculated according
to Eq. (\protect\ref{res}). The solid and long--dashed 
curves are for the potentials
with the parameters 
(\protect\ref{par}) and (\protect\ref{p2}), respectively.}
\end{figure}

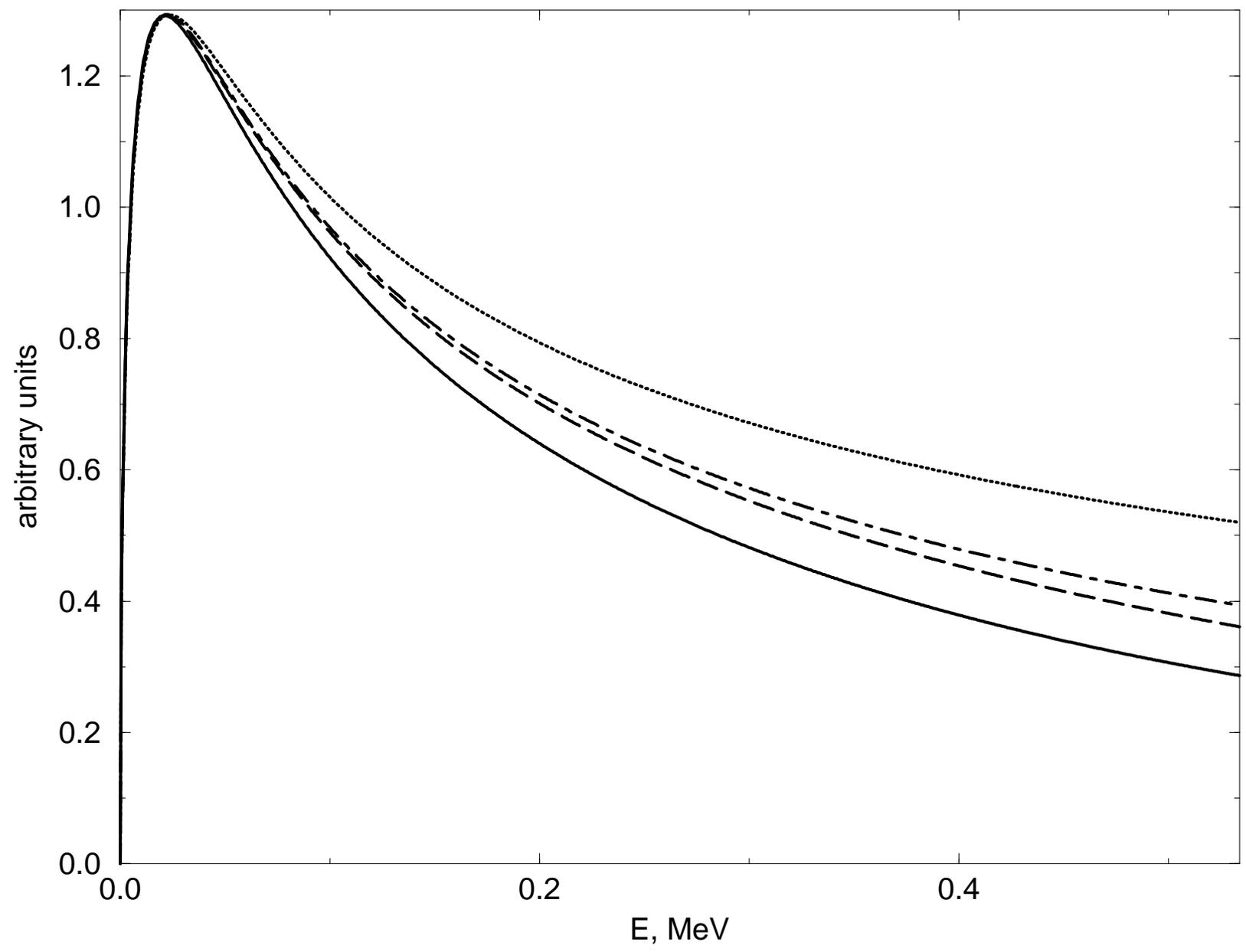
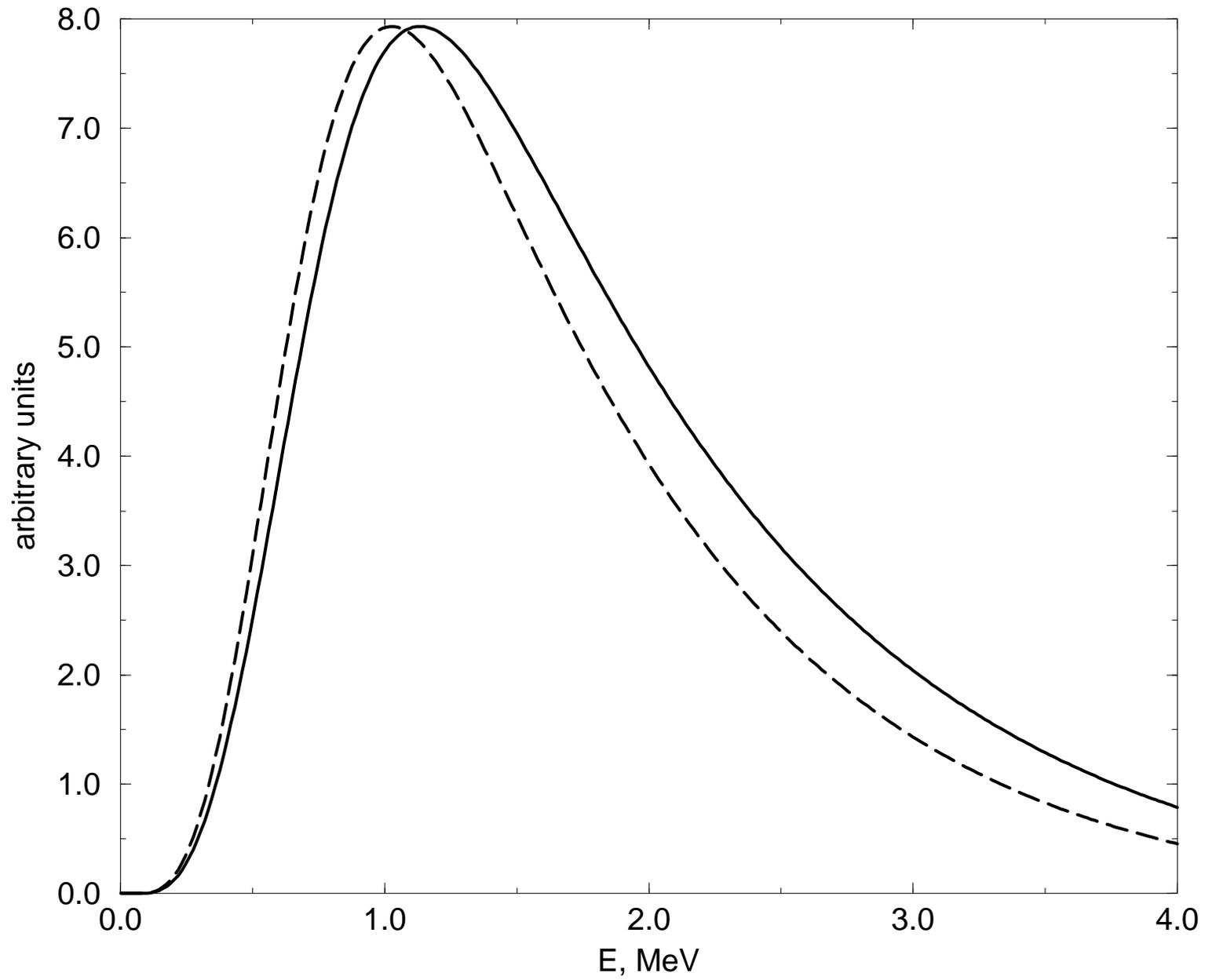
\end{document}